\documentclass[sigconf,screen]{acmart}
\acmBooktitle{Companion Proceedings of the 33rd ACM Symposium on the Foundations of Software Engineering (FSE '25), June 23--27, 2025, Trondheim, Norway}
\AtBeginDocument{%
  }

\usepackage{shortcuts}
\usepackage{array}
\usepackage{multirow}
\usepackage{tabularx}
\usepackage{pifont}
\usepackage{booktabs}

\begin{document}

\title{Tracking GPTs Third Party Service: Automation, Analysis, and Insights}

\author{Chuan Yan}
\orcid{0000-0003-4855-1912}
\affiliation{%
  \institution{University of Queensland}
  \city{Brisbane}
  \state{QLD}
  \country{Australia}
}

\author{Liuhuo Wan}
\orcid{0009-0004-7090-1493}
\affiliation{%
   \institution{University of Queensland}
   \city{Brisbane}
  \state{QLD}
   \country{Australia}
}

\author{Bowei Guan}
\orcid{0009-0007-1414-8523}
\affiliation{%
  \institution{University of Queensland}
  \city{Brisbane}
  \state{QLD}
  \country{Australia}
}

\author{Fengqi Yu}
\orcid{0009-0007-8162-1982}
\affiliation{%
  \institution{University of Queensland}
  \city{Brisbane}
  \state{QLD}
  \country{Australia}
}

\author{Guangdong Bai}
\orcid{0000-0002-6390-9890}
\affiliation{%
  \institution{National University of Singapore}
  \country{}
}
\affiliation{%
  \institution{University of Queensland}
  \state{QLD}
  \country{Australia}
}

\author{Jin Song Dong}
\orcid{0000-0002-6512-8326}
\affiliation{%
  \institution{National University of Singapore}
  \country{}
}



\begin{abstract}
ChatGPT has quickly advanced from simple natural language processing to tackling more sophisticated and specialized tasks. Drawing inspiration from the success of mobile app ecosystems, OpenAI allows developers to create applications that interact with third-party services, known as GPTs. 
GPTs can choose to leverage third-party services to integrate with specialized APIs for domain-specific applications. However, the way these disclose privacy setting information limits accessibility and analysis, making it challenging to systematically evaluate the data privacy implications of third-party integrate to GPTs.
In order to support academic research on the integration of third-party services in GPTs, we introduce \tool, an automated framework designed to extract GPTs' privacy settings. \tool provides academic researchers with real-time, reliable metadata on third-party services used by GPTs, enabling in-depth analysis of their integration, compliance, and potential security risks. By systematically collecting and structuring this data, \tool facilitates large-scale research on the transparency and regulatory challenges associated with the GPT app ecosystem.
\end{abstract}




\maketitle
\section{Introduction}
ChatGPT, introduced by OpenAI~\cite{openai2023creating} in 2023, is a leading large language model~(LLM) that showcases cutting-edge advancements in AI-powered natural language processing~(NLP) technology. It has been widely adopted across various domains, utilizing GPT models to streamline tasks and improve efficiency~\cite{yan2024investigating,yuan2024evaluating,zhang2025unveiling,li2024drowzee,wang2024corelocker, wan2024don}. As LLMs continue to evolve, user demand for personalized and customizable AI solutions has been steadily increasing. In response this trend, OpenAI introduce the GPT Store~\cite{openai2024data}, featuring GPT applications~(GPTs) that not only enable access to real-time data, perform computations, and integrate with third-party services but also allow users to create and customize GPTs without any coding. This innovation marks a significant step toward a more open and diverse GPT ecosystem.

However, while interactions with third-party services enhance the functionality of GPTs, they also introduce significant security risks. Previous studies analyze and confirm these security concerns from multiple perspective. For example, attackers can exploit leaked GPTs data to clone fraudulent GPTs for phishing or scams~\cite{xie2024llm,iqbal2024llm}. Certain third-party services are vulnerable to unauthorized access, leading to potential privacy breaches~\cite{yan2024exploring}. And uploaded knowledge files could be inadvertently exposed, increasing the likelihood of data leaks~\cite{yan2025understanding}. To conduct these security analyses, it is essential to first obtain the third-party service information provided by GPTs developers. Most existing research relies on foundational GPTs data collected by third-party platforms. However, third-party platforms suffer from poor data timeliness, making it difficult to track updates to GPTs. Researchers often face challenges in obtaining up-to-date information, which limits the accuracy and reliability of analyses. For example, in its latest version, AI PDF Drive~\cite{aipdf2024myaidrive} specifies the domain names and corresponding privacy policies of three third-party services~(\texttt{account.myaidrive.com}, \texttt{aipdf.myaidrive.com},\texttt{pdf-creator.myaidrive.com}). Nevertheless, GPTs App~\cite{gptsapp2024pdfai} only records two domains of them, preventing researchers from obtaining a comprehensive view of its most recent privacy compliance updates.
\paragraph{Our work}
To reduce experimental costs while improving data timeliness, we develop GPTs-ThirdSpy, a framework for automatically extracting GPTs third-party privacy setting data from the GPT Store. A major challenge is that OpenAI has implemented strict anti-scraping mechanisms~\cite{openai24verify} and dynamic content loading in the GPT Store, making traditional automation tools such as Selenium~\cite{Selenium} and Puppeteer~\cite{Puppeteer} ineffective for interacting with its page structure. To overcome this, we build on previous research~\cite{yan2024exploring} by leveraging AppleScript~\cite{applescript} to simulate user interactions, bypassing the anti-scraping defenses. Additionally, we introduce an innovative approach using cliclick~\cite{cliclick}, enabling precise position-based clicking to extract the complete GPTs privacy setting data.
\tool detects 109 GPTs relying on third-party services within a sample of 500 GPTs and successfully extracts their associated privacy setting data. In future work, we will build a more comprehensive dataset that covers a broader range of GPTs. This enables future research by supporting privacy compliance audits, supporting privacy compliance audits, penetration testing, and other security analyses to comprehensively examine the usage patterns and potential risks of third-party services in GPTs.
\paragraph{Contribution}
The main contributions of this work are as follows.

\begin{itemize}
    \item \textbf{Metadata-driven categorization of GPTs.}
    We categorize GPTs into three distinct types based on their metadata characteristics, providing a structured framework to better understand their functional differences and usage patterns. 
    
    \item \textbf{A systematic security assessment tool.}
    We propose \tool, a framework designed to automatically detect and extract GPTs third-party service data from the GPT Store. Unlike existing methods that rely on third-party platforms, our framework captures real-time and accurate data, providing a more reliable foundation for future research. 
    
    \item \textbf{Analyzing Third-Party Service Usage in GPT Store and Future Directions.} 
    Our results reveal that the \emph{status quo} of utilizing third-party services in GPTs. This not only provides valuable data for privacy compliance assessments but also supports future research in security evaluations, malicious GPTs detection, and other areas that rely on up-to-date data.
    
\end{itemize}

\paragraph{Availability} The source code of \tool and relevant artifacts are available on Github~\cite{github}. 
\section{Experimental Study} \label{sec:es}
To explore the deployment process and regulations of third-party services on the \gs, we specifically develop GPTs as a test case for validation.
Table~\ref{tab:metadata} presents the seven types of metadata defined by OpenAI for GPTs. Among them, \emph{name} and \emph{instructions} are mandatory, meaning that every GPTs must include these two metadata types (our work does not consider empty-shell GPTs that only include \emph{name}). The remaining five metadata types are optional, allowing GPTs to incorporate them as needed to enhance functionally and personalization. We use $g$ represents a GPTs, $\mathcal{G}$ represents GPTs set,  thus we have $\mathcal{G} = \{g |\,g : \{name, instructions, description', conv\\ersation', knowledge', capabilities', actions'\} \}$, where $'$ denotes optional metadata.

Based on the functions and properties of different GPTs metadata, we define three types of GPTs.
\\

\paragraph{Prompt-based GPTs ($\mathcal{G}_p$)}
\emph{Prompt-based indicates that the GPTs relies solely on prompts for description and generation, without incorporating any background knowledge or external resources.}
\begin{equation}
\label{equ:1}
    g \in \mathcal{G}_p \Leftrightarrow \{knowledge\} \notin g \land \{actions\} \notin g
\end{equation}

\paragraph{Knowledge-based GPTs ($\mathcal{G}_k$)}
\emph{Knowledge-based GPTs allow developers to upload local knowledge, such as text documents. By integrating external information, these GPTs enhance their accuracy, retain domain-specific expertise, and generate more context-aware responses.}
\begin{equation}
\label{equ:2}
    g \in \mathcal{G}_k \Leftrightarrow \{knowledge\} \in g \land \{actions\} \notin g
\end{equation}

\paragraph{Action-based GPTs ($\mathcal{G}_a$)}
\emph{Action-based GPTs can interact with external systems by making API calls, enabling dynamic and real-time responses. Unlike prompt-based or knowledge-based, these GPTs can fetch live data, execute commands, and integrate with third-party services.}
\begin{equation}
\label{equ:3}
    g \in \mathcal{G}_a \Leftrightarrow  \{actions\} \in g
\end{equation}

Among the three types of GPTs, Prompt-based GPTs are the most fundamental, pioneering a zero-code approach to creating custom applications, allowing users to build their own GPTs simply by writing prompts. A step further is Knowledge-based GPTs, which enable developers to upload local knowledge, equipping GPTs with domain-specific expertise for more accurate responses and personalized services. The only type that interacts with third-party services is Action-based GPTs, which leverage APIs to access external data and perform complex computations. However, this reliance on external services also introduces greater security and privacy risks, including data breaches and unauthorized API access~\cite{yan2024exploring,hou2024security,xie2024llm}. Therefore, research on this category of GPTs is essential to address these challenges and develop more secure and compliant management frameworks.

\begin{table}[t]
\centering
\renewcommand{\arraystretch}{1.2} 
\setlength{\tabcolsep}{3pt} 
\caption{Metadata of GPTs\label{tab:metadata}}

\begin{tabularx}{\columnwidth}{p{0.21\columnwidth}|p{0.75\columnwidth}} 
\hline

\hline \textbf{Metadata}  & \textbf{Explanation} \\ 
\hline
Name$\dagger$ & The non-unique, potentially repetitive identifier for GPTs. \\ 
Description$\ddagger$  & A brief summary of what the GPTs does. \\ 
Instructions$\dagger$  & Define the behavior and limitations. \\ 
Conversation$\ddagger$  & Predefined prompts to guide user interactions. \\ 
Knowledge$\ddagger$  & Uploaded files that enhance the GPTs' responses. \\ 
Capabilities$\ddagger$  & Additional features the GPTs can use. \\ 
Actions$\ddagger$  & Custom third-party API integrations for external interactions. \\ \hline

\hline
\end{tabularx}
\begin{flushleft}
   \small
    \textsuperscript{$\dagger$} Mandatory: Metadata that developers are required to provide on a mandatory basis.
    
    \textsuperscript{$\ddagger$} Discretionary: Metadata that developers may choose to provide.

\end{flushleft}
\end{table}

\section{GPTs-ThirdSpy}
\begin{figure*}[ht]
    \centering
    \includegraphics[width=0.95\textwidth]{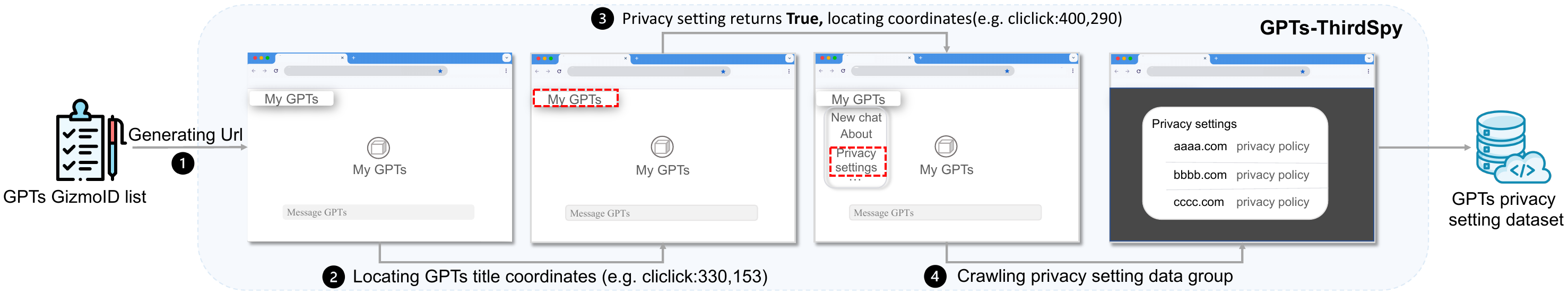}
    \caption{The workflow of GPTs-ThirdSpy}
    \label{fig:framework}
\end{figure*}
Leveraging prior knowledge from the experimental study in Section~\ref{sec:es}, we design \tool, which can automatically crawl third-party service data from GPTs. Figure~\ref{fig:framework} illustrates the complete workflow of \tool.
\paragraph{Workflow}
Each GPTs has a GizmoID, which serves as its unique 9-character alphanumeric identifier and provides access to its interaction page. Based on this, we first construct the access link based on the GizmoID of each GPTs~\ding{182}~(e.g., GizmoID: 1abcD2EFG, Url: https://chatgpt.com/g/g-1abcD2EFG).
After successfully accessing the GPTs interaction page, the next step it to retrieve its detailed information, which will only be shown after clicking the dropdown button. Therefore, we need to locate and click this button. Since the GPTs interaction page has strict anti-automation mechanisms, methods relying on element IDs for interaction, such as JavaScript or Selenium, often fail to execute or trigger CAPTCHA verification, making automated testing ineffective. 
To overcome this challenge, we use AppleScript in combination with Cliclick. Cliclick is a commandline tool for macOS that simulates mouse clicks. By precomputing the absolute screen coordinates of the GPTs dropdown button, we can bypass anti-automation restrictions and successfully expand the GPTs details panel~\ding{183}.

According to OpenAI's guidelines, if a GPTs utilizes external APIs~(i.e. Actions), it must provide a corresponding privacy policy. As a result, a ``Privacy settings'' button appears in the GPTs's dropdown menu. Leveraging this characteristic, we precompute the absolute screen coordinate of the ``Privacy settings'' button and use Cliclick for precise clicking, ensuring seamless access to the relevant privacy data~\ding{184}.
When the ``Privacy settings'' button is clicked, a window pops up displaying the domain names of all third-party services used by the GPTs, along with their corresponding privacy policy links. To comprehensively collect this data, we iterate through each domain entry, extract and parse its associated privacy policy link, and access these links to crawl the underlying privacy policy text. Finally, we organize the collected data into the GPTs privacy setting dataset for further analysis and research~\ding{185}.

\section{Dataset}
\paragraph{Data source and experiment setup}
We use the 500 most popular GPTs~(ranked by conversation count) provided by GPTsHunter as our data source. Since \tool is built with AppleScript, it relies on the macOS environment. Therefore, we deploy it on two Mac devices: a 16GB M1 Pro and a 16GB M2 Pro.

\subsection{Domain Analysis}
\begin{figure}[t]
    \centering
    \includegraphics[width=0.47\textwidth]{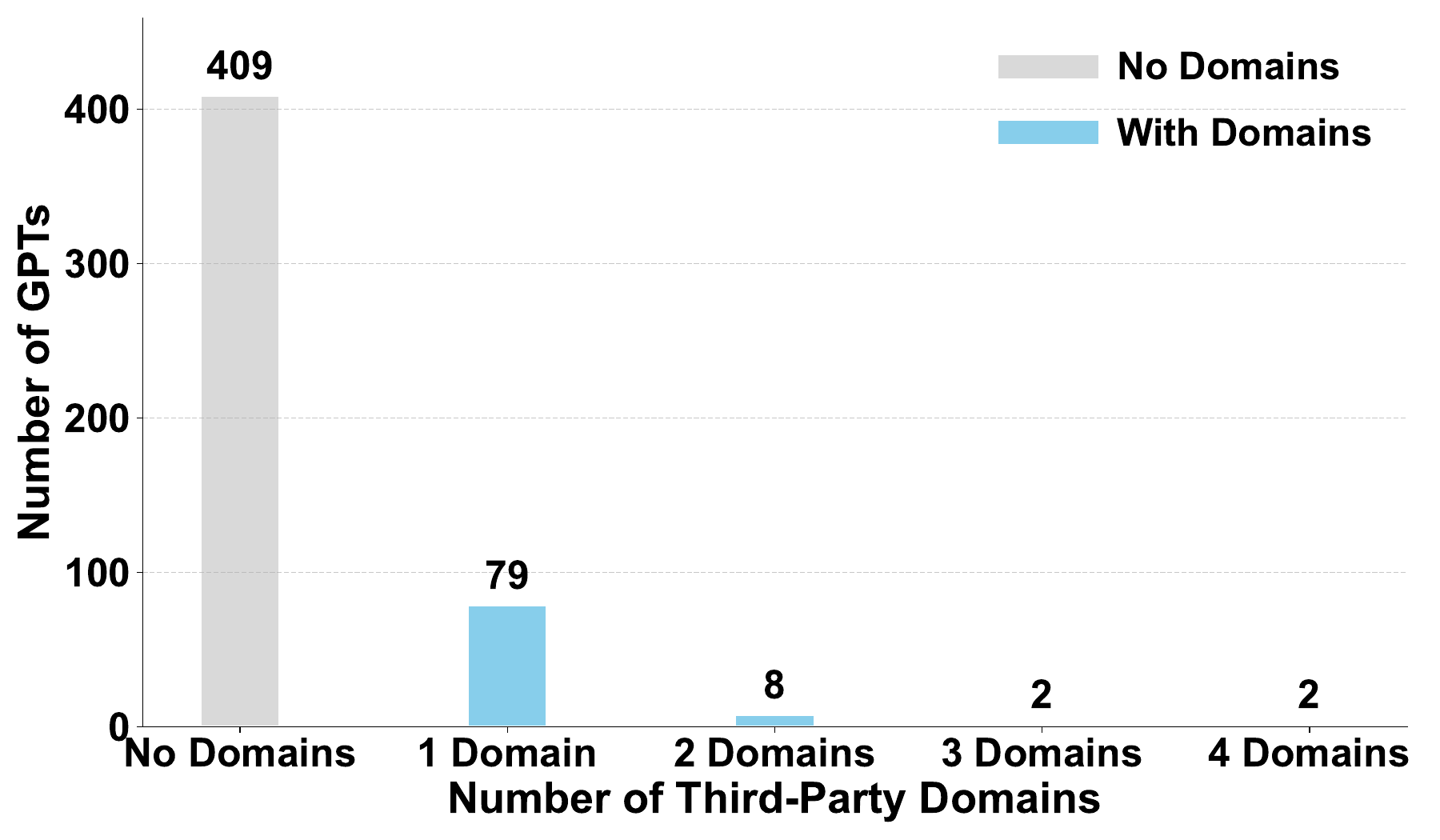}
    \caption{GPTs domain count distribution}
    \label{fig:domain_name}

\end{figure}

Figure~\ref{fig:domain_name} illustrates the distribution of third-party service usage among the 500 most popular GPTs. Among them, 409 GPTs do not rely on any third-party services. These GPTs operate solely based on their own knowledge or GPT model, without calling external API or transmitting additional data, making them more self-contained in terms of privacy protection.
79 GPTs integrate a single third-party service, primarily to extend their functionality, such as accessing external databases, providing real-time information, or enhancing computational capabilities. 
This represents the most common pattern of third-party service usage among GPTs.
Additionally, 12 GPTs utilize two or more third-party services. Manual verification reveals that these GPTs often require more advanced functionalities, such as cross-platform integrations, real-time data analysis, or leveraging multiple AI APIs for task optimization. However, this also introduces greater security and privacy risks, as data may be processed across multiple external systems, increasing the complexity of access control and the potential for information leakage. We leave the analysis and assessment of these security and privacy risks for future work.

\begin{figure}[t]
    \centering
    \includegraphics[width=0.49\textwidth]{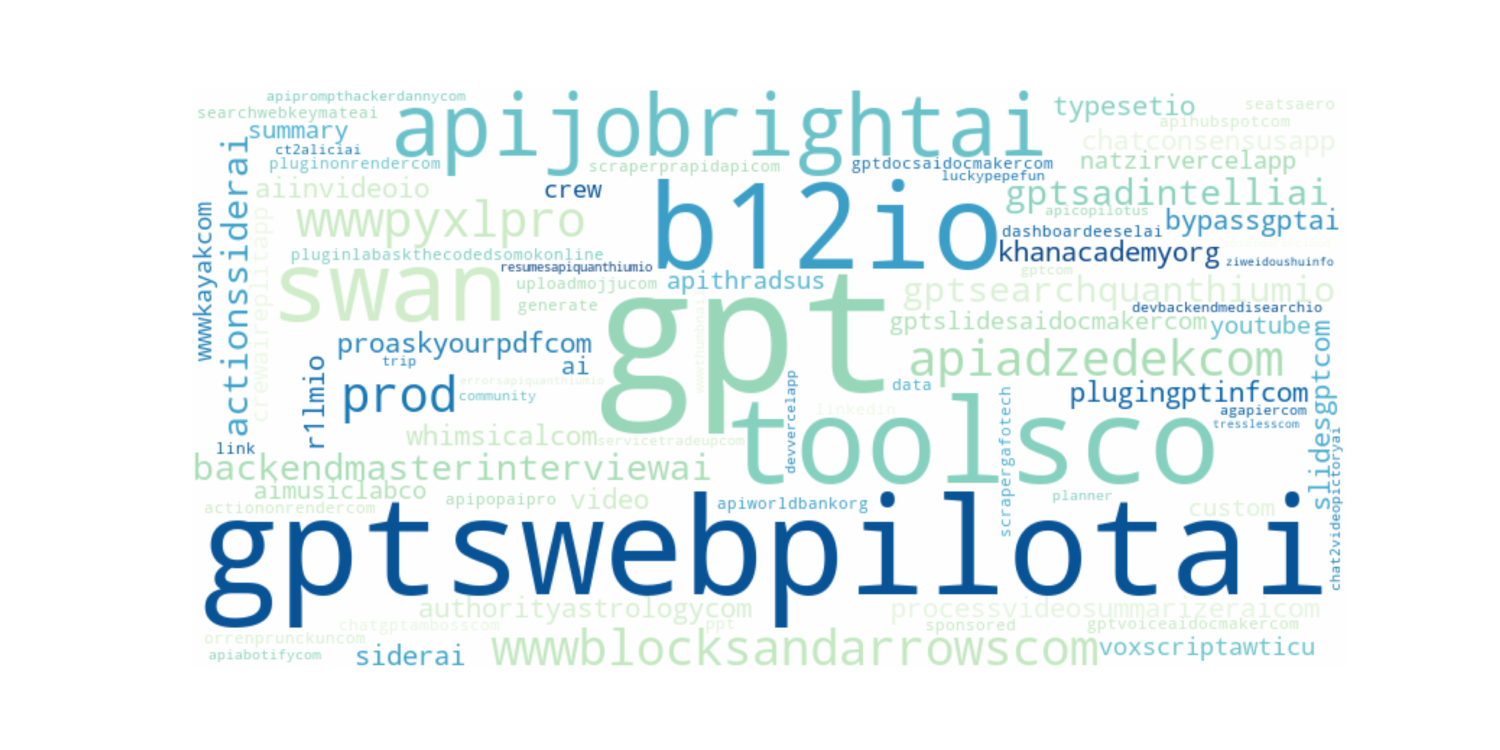}
    \caption{Domain Usage Distribution of GPTs}
    \label{fig:wordcloud}

\end{figure}

Figure~\ref{fig:wordcloud} shows the distribution of these third-party service domains. Third-party services for AI application development are the most frequently used, with services such as \texttt{gpts.WebPilot.ai}~\cite{webpilot}, \texttt{gpt-tools.co}~\cite{gptbuilder} and \texttt{b12.io}~\cite{b12}. Following closely behind are third-party services that leverage AI for specialized tasks. For example, \texttt{swan-api.jobright.ai}~\cite{jobright} uses AI to assist users in job searching, improving efficiency and optimizing the matching process.

\subsection{Privacy Policy}
We systematically access and store the text content of privacy policy links in our dataset for further analysis. Although OpenAI mandates that GPTs using third-party services provide corresponding privacy policy links, it does not enforce content review, leading to potential issues with validity and compliance. Among the 109 domains associated with third-party services, 92 privacy policy links are accessible, while others exhibit various issues. 9 links are invalid, including 1 GPTs that directly uses a placeholder link~(i.e. \texttt{https://app.example.com/privacy\_policy}). 5 GPTs provide only the homepage of the third-party service instead of a dedicated privacy policy, which could prevent users from obtaining clear information on data handling and privacy protection. Additionally, 2 links result in connection timeouts, while another 2 return server errors. These findings indicate that despite OpenAI's requirement for privacy policy links, many GPTs still fail to provide valid or compliant policies, underscoring the need for stricter review and regulatory mechanisms to ensure users have access to meaningful full and enforceable privacy data.
\begin{table}[t]
\
\caption{Privacy policy accessibility distribution\label{tab:pp_dis}}
\resizebox{0.95\linewidth}{!}{%
\begin{tabular}{c|c c c c}
\hline

\hline   \multicolumn{1}{c|}{ \textbf{Accessible} } & \multicolumn{4}{c}{ \textbf{Inaccessible} }\\ \cline{1-5}

  \multicolumn{1}{c|}{-} & \multicolumn{1}{c|}{Broken link} & \multicolumn{1}{c|}{Official website} & \multicolumn{1}{c|}{Timeout} & \multicolumn{1}{c}{Server error} \\ \hline
 
 92 &\multicolumn{1}{c|}{9} &\multicolumn{1}{c|}{5} &\multicolumn{1}{c|}{2} &\multicolumn{1}{c}{2} \\ \hline
 
 \hline

\end{tabular}}
\end{table}

\section{Discussion}
\tool addresses the issue of outdated GPTs metadata, enabling developers to access the latest third-party service data directly from the official GPT Store in real-time, ensuring the accuracy and reliability of research and applications. Based on these features, \tool can be further extended in several key areas.

\paragraph{Privacy compliance}
Academic researchers and industry regulators can use this tool to monitor the enforcement of GPTs' privacy policy in real-time, identifying non-compliant third-party data-sharing practices and promoting a more transparent and secure AI ecosystem, building upon prior work~\cite{xie2022scrutinizing, yan2024quality}. Additionally, \tool can be extended to automatically detect inconsistencies in privacy policies, analyzing whether GPTs provide contradictory privacy statements across different periods or platforms.

\paragraph{Third-Party service monitoring}
\tool also can be used to track changes in third-party services that GPTs rely on, allowing it to detect when a GPTs adds or replaces an external API. In such cases, the system can issue alerts, prompting regulatory agencies or enterprises to reassess data access permissions. This functionality is particularly valuable for identifying sudden shifts in data-sharing practices, ensuring that new integrations comply with privacy and security standards, and preventing unauthorized access to sensitive user information. Moreover, by maintaining a historical record of these changes, \tool enables trend analysis, helping researchers and policymakers understand long-term patterns in GPTs' reliance on third-party services and anticipate potential regulatory challenges.

\paragraph{Penetration testing}
\tool can be further extended to conduct penetration testing on third-party service domains relied upon by GPTs, allowing for security assessments and the identification of potential vulnerabilities. Since third-party APIs used by GPTs could serve as entry points for data breaches or supply chain attacks~\cite{wan2024safe}, the tool can automatically identify and track these domains while integrating penetration testing techniques for comprehensive security evaluation. \tool also monitors third-party services for risks such as API abuse, authentication flaws, and potential privilege escalation vulnerabilities, enabling enterprises, researchers, and regulatory agencies to assess the security of the GPT app ecosystem more effectively and proactively mitigate emerging threats.
\section{Conclusion}
In this work, we first categorize GPTs into three distinct types based on their metadata characteristics. This categorization not only enhances understanding of GPTs in terms of functionality, dependencies, and privacy compliance but also establishes a systematic foundation for future research. Additionally, we develop \tool, an automated framework for extracting third-party service data from GPTs. This framework enables real-time and precise retrieval of GPTs' interactions with external services, providing researchers with a reliable analytical foundation. 

\begin{acks}
This research has been partially supported by Australian Research Council Discovery Projects (DP230101196, DP240103068) and the Ministry of Education, Singapore under its Academic Research Fund Tier 3 (MOET32020-0003). 
\end{acks}

\bibliographystyle{ACM-Reference-Format}

\bibliography{paper}

\end{document}